\documentclass[twocolumn,showpacs,preprintnumbers,amsmath,amssymb,aps,prb,
superscriptaddress
]{revtex4}
\usepackage{graphicx}
\usepackage[latin1]{inputenc}

\def\tlse{Laboratoire de Physique Th\'eorique, Universit\'e Paul Sabatier, CNRS, 
118 Route de Narbonne, 31400 Toulouse, France.}
\begin{document}

\title{Confinement vs Deconfinement of Cooper Pairs in One-Dimensional
Spin-3/2 Fermionic Cold Atoms}

\author{S.\ Capponi}
\email{capponi@irsamc.ups-tlse.fr} \affiliation{\tlse}
\author{G.\ Roux} \affiliation{\tlse}
\author{P. Azaria} \affiliation{Laboratoire de Physique Th\'eorique de la Matiere
Condens\'ee, Universit\'e Pierre et Marie Curie, CNRS, 4 Place
Jussieu, 75005 Paris, France.}
\author{E. Boulat} \affiliation{Laboratoire MPQ, Universit\'e Paris 7, CNRS, 2 Place Jussieu,
75005 Paris, France.}
\author{P. Lecheminant} \affiliation{Laboratoire de Physique Th\'eorique et Mod\'elisation,
Universit\'e de Cergy-Pontoise, CNRS, 2 Avenue Adolphe Chauvin, 95302
Cergy-Pontoise, France.}

\date{\today}
\pacs{{67.90.+z}, {05.30.Fk}, {03.75.Ss}, {71.10.Fd}
}

\begin{abstract}
The phase diagram of spin-3/2 fermionic cold atoms trapped in a
one-dimensional optical lattice is investigated at quarter filling
(one atom per site) by means of large-scale numerical simulations.  In
full agreement with a recent low-energy approach, we find two phases
with confined and deconfined Cooper pairs separated by an Ising
quantum phase transition.  The leading instability in the confined
phase is an atomic-density wave with subdominant quartet superfluid
instability made of four fermions. Finally, we reveal the existence of a bond-ordered Mott 
insulating phase in some part of the repulsive regime.
\end{abstract}

\maketitle
Loading cold atomic gases into optical lattice allows for the
realization of bosonic and fermionic lattice models and the
experimental study of exotic quantum phases~\cite{review}. Ultracold
atomic systems also offer an opportunity to investigate the effect of
spin degeneracy since the atomic total angular momentum $F$ can be
larger than $1/2$ resulting in $2F+1$ hyperfine states.  This
high-spin physics is expected to stabilize novel exotic phases.  In
this respect, various superfluid condensates, Mott insulating phases,
and interesting vortex structures have been found in spinor bosonic
atoms with $F \ge 1$ \cite{ho,boson}. These theoretical predictions
might be checked in the context of Bose-Einstein condensates of
sodium, rubidium atoms and in spin-3 atom of $^{52}$Cr~\cite{spin3}.
The spin-degeneracy in fermionic atoms is also expected to give rise
to some interesting superfluid and Mott phases.  In particular, a
molecular superfluid phase might be stabilized where more than two
fermions form a bound state.  Though such non-trivial superfluid
behavior has been previously found in different contexts
\cite{nozieres}, it has been advocated recently that the formation of
bound-state of Cooper pairs is likely to occur in general half-integer
$F>1/2$ ultracold atomic fermionic systems
\cite{Lecheminant2005,Wu2005,miyake}.  In the spin $F=3/2$ case, it
has been predicted on the basis of a low-energy study
\cite{Wu2005,Lecheminant2005} in one dimension that a quartetting
superfluid phase, i.e. a bound-state of two Cooper pairs, might be
stabilized by strong enough attractive interactions.  The simplest
lattice Hamiltonian to describe spin-3/2 atoms with s-wave scattering
interactions in 1D optical lattice takes the form of a Hubbard-like
model \cite{ho}:
\begin{eqnarray}
{\cal H}
&=& -t \sum_{i,\alpha} \left[c^{\dagger}_{\alpha,i}
c_{\alpha, i+1} + {\rm H.c.} \right]
\nonumber \\
& &+ U_0 \sum_{i} 
P_{00,i}^{\dagger} P_{00,i}
+ U_2 \sum_{i,m}
P_{2m,i}^{\dagger} P_{2m,i},
\label{hubbardSgen}
\end{eqnarray}
where $c^{\dagger}_{\alpha,i}$ is the fermion creation operator
corresponding to the four hyperfine states $\alpha = \pm 1/2, \pm
3/2$.  The singlet and quintet pairing operators in
Eq.~(\ref{hubbardSgen}) are defined through the Clebsch-Gordan
coefficient for two indistinguishable particles: $P^{\dagger}_{JM,i} =
\sum_{\alpha \beta} \langle JM|F,F;\alpha \beta\rangle
c^{\dagger}_{\alpha,i}c^{\dagger}_{\beta,i}$.  As it appears, it is
more enlightening to express model (\ref{hubbardSgen}) in terms of the
density ($n_i = \sum_{\alpha} c^{\dagger}_{\alpha,i} c_{\alpha,i}$)
and the singlet pairing operator
($P^{\dagger}_{00,i}=P^{\dagger}_{i}=\frac{1}{\sqrt{2}}[c^{\dagger}_{\frac
3 2,i}c^{\dagger}_{-\frac 3 2,i} - c^{\dagger}_{\frac 1
2,i}c^{\dagger}_{-\frac 1 2,i}]$):
\begin{eqnarray}
{\cal H}
&=& -t \sum_{i,\alpha} [c^{\dagger}_{\alpha,i} c_{\alpha,i+1} 
+ {\rm H.c.} ]
\nonumber \\
& & +\frac{U}{2} \sum_i n_i^2 + V \sum_i
P_{i}^{\dagger} P_{i},
\label{hubbardS}
\end{eqnarray}
with $U= 2 U_2$ and $V = U_0 - U_2$. Model~(\ref{hubbardS})
generically displays an exact SO(5) extended spin symmetry and an
SU(4) symmetry in the particular case $U_0=U_2$, i.e. $V=0$
\cite{zhang}. In sharp contrast with the spin $F = 1/2$ case where
both interacting terms in Eq.~(\ref{hubbardS}) are proportional, these
terms are independent for $F=3/2$ and strongly compete. While the
$V$-term favors the pairing of two fermions for negative $V$, an
attractive $U$-interaction might favor the formation of a quartet $Q_i
= c_{-\frac 3 2,i} c_{-\frac 1 2,i} c_{\frac 1 2,i} c_{\frac 3
2,i}$. In fact, it has been recognized in
Ref.~\onlinecite{Lecheminant2005} that the above competition reveals
itself through a non-trivial discrete symmetry of the problem. Indeed,
model~(\ref{hubbardS}) possesses, on top of the SO(5) symmetry, a
${\mathbb{Z}}_2$ discrete symmetry ${\cal U}$: $c_{\alpha,i}
\rightarrow e^{i\pi/2} c_{\alpha,i}$ which plays a crucial role in the
low-energy physics since, as $P_{i}$ is odd under ${\cal U}$, the
formation of a quasi-long range BCS phase requires ${\cal U}$ to be
spontaneously broken. When ${\cal U}$ is unbroken the BCS instability
is strongly suppressed and the leading superfluid instability is made
of four fermions i.e. a quartet which is even under ${\cal U}$.
Such a two-phase structure has been  recently predicted away from
half-filling in the weak coupling limit by means of a low-energy
approach~\cite{Wu2005,Lecheminant2005,controzzi}.
\begin{figure}[t]
\includegraphics[width=8.4cm,clip]{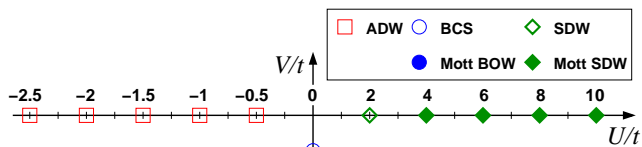}
\caption{(Color online) Phase diagram of the spin-3/2 Hubbard chain
(\ref{hubbardS}) at quarter filling from QMC and DMRG calculations for
$V\leq 0$ (see text for definitions).}
\label{phase_diagram.fig}
\end{figure}

In this letter, we investigate numerically the phase diagram of
model~(\ref{hubbardS}) for $V \le 0$ at quarter filling (one atom per
site) by means of quantum Monte-Carlo (QMC) and Density-Matrix
Renormalization Group~\cite{dmrg} (DMRG) simulations. Physical
properties are investigated by computing the one-particle,
density, pairing as well as quartetting correlation functions,
respectively: $G(x)=\langle c_{\alpha,i}^{\dag} c_{\alpha,i + x}
\rangle$, $N(x) = \langle n_i n_{i+x} \rangle$, $P(x)=\langle P_i
P^{\dag}_{i+x} \rangle$ and $Q(x) = \langle Q_i Q^{\dag}_{i+x}
\rangle$.  For the QMC simulations, we used the projector auxiliary
field QMC algorithm (see Ref.~\onlinecite{Assaad1997} for the details
of the algorithm) in the regime $V\le 0 \quad {\rm and} \quad U\le -
3V/4$ where the fermionic algorithm has no sign
problem~\cite{zhang}. We have studied \emph{periodic} chains with
linear size up to $L=180$ with a typical projection parameter $\Theta
t =10$ and a Trotter time slice $\Delta t = 0.05$. Most of DMRG
calculations were performed on \emph{open} chains with $L=60$
sites and keeping $M=1400$ states~\cite{note}. The
resulting phase diagram at quarter filling is presented in
Fig.~\ref{phase_diagram.fig} and we now turn to the discussion of the
physical properties of the different phases.
 \begin{figure}[t]
\includegraphics[height=5.2cm,clip]{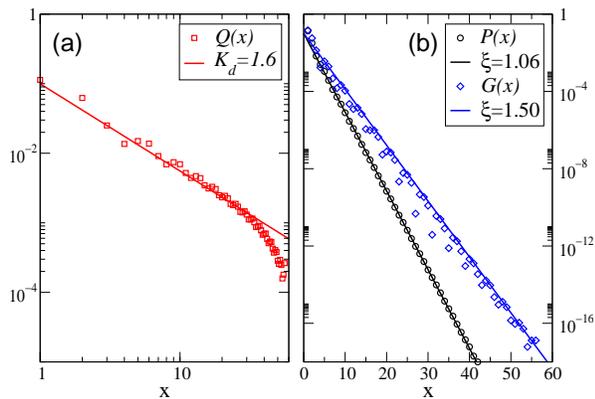}
\caption{(Color online) Correlation functions for $U=-1.5 t$ and $V=0$
obtained by DMRG. {\bf(a)} Power-law behavior of the quartet
correlations parametrized by Luttinger exponent $K_d$. {\bf(b)}
Short-range behavior of the one-particle Green function and pairing
correlations ($\xi$ denotes the correlation lengths).}
\label{su4.fig}
\end{figure}
\begin{figure}[t]
\includegraphics[height=5.2cm,clip]{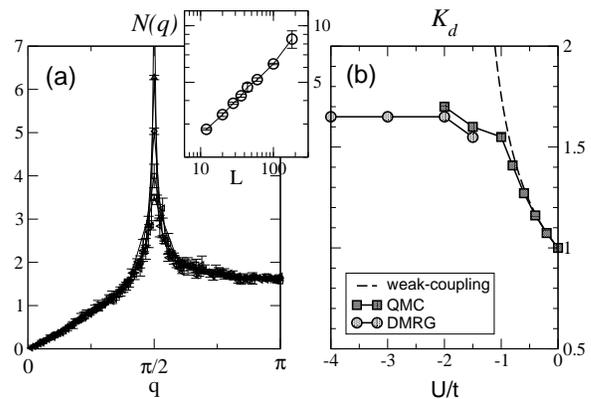}
\caption{{\bf(a)} Fourier transform $N(q)$ of the density
correlations obtained from QMC ($U=-t$ and $V=0$). The linear
dispersion at small $q$ gives access to the Luttinger parameter
$K_d$. Insert~: the scaling of the peak at $2k_F$ vs $L$ signals an
ADW phase. {\bf (b)} Luttinger exponent $K_d$ as a function as $U$. }
\label{Kd_on_SU4.fig}
\end{figure}
\begin{figure}[t]
\includegraphics[height=5.2cm,clip]{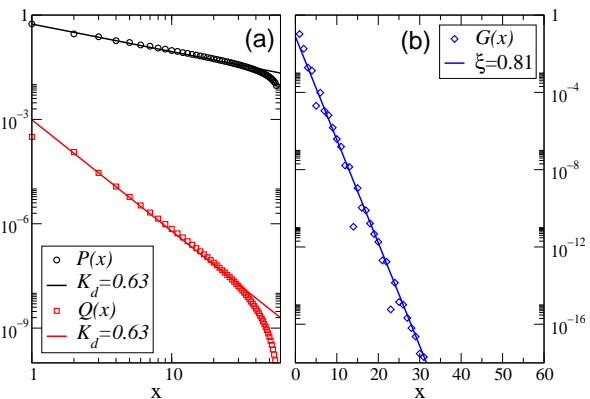}
\caption{(Color online) Correlation functions for $U=0$ and $V=-3 t$
from DMRG. {\bf(a)} Pair and quartet correlations are algebraic with
$Q(x) \sim P(x)^4$. {\bf(b)} Short-range behavior of the one-particle
Green function.}
\label{BCS.fig}
\end{figure}
\begin{figure}[t]
\includegraphics[height=5.2cm,clip]{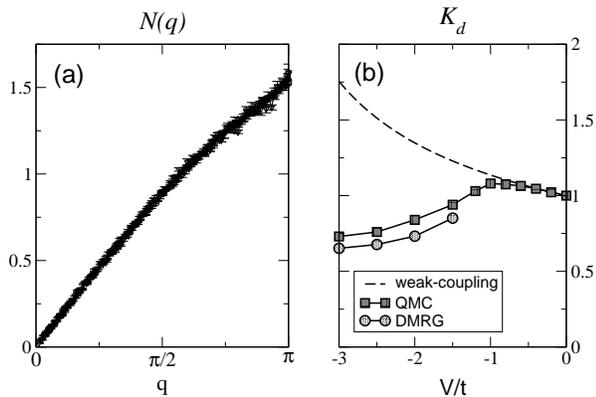}
\caption{{\bf (a)} Fourier transform of the density correlations
obtained by QMC for $U=0$ and $V=1.5t$. {\bf (b)} The Luttinger
parameter $K_d$ as a function of $V$ when $U=0$.}
\label{Kd_on_BCS.fig}
\end{figure}

{\it Confined Phase --} The phase with $U <0$ and small $|V|$ is
characterized by the existence of a spin gap and an unbroken
${\mathbb{Z}}_2$ discrete symmetry ${\cal U}$ which marks the onset of
an Atomic Density Wave (ADW) and quartet superfluid quasi-long-range
orderings. Indeed, for the typical value of $U=-1.5t$ and $V=0$, we
observe in Fig.~\ref{su4.fig}(b) that both the pairing correlations
and the one-particle Green function decay exponentially with 
distance. In contrast, the quartet correlations are algebraic as it
can be seen in Fig.~\ref{su4.fig}(a). We have checked by a direct
evaluation that the four-particle gap vanishes. We can thus deduce
that the short-range character of $P(x)$ is due to the confinement of
Cooper pairs which stems from the unbroken ${\cal U}$ symmetry.  The
above results extend in the whole confined phase (squares in
Fig.~\ref{phase_diagram.fig}). In this phase, the superfluid
instability is of a molecular type made of four fermions: a quartet.
However, the density correlations also display a power-law behavior
with dominant oscillations at $2k_F = \pi/2$ as it is clearly seen
from the Fourier transform $N(q)$ of $N(x)$ presented in
Fig.~\ref{Kd_on_SU4.fig}(a). The question that naturally arises is
which instability dominates in this phase. The answer depends on the
value of the non-universal Luttinger parameter $K_d$ which stems from
the critical behavior of the density degrees of freedom. Indeed, the
quartet and $2k_F$-ADW equal-time correlation functions have been
found in Refs. \cite{Wu2005,Lecheminant2005} to behave at long
distance as $Q(x)\sim x^{-2/K_d}$ and $N(x)\sim \cos(\pi x/2)
x^{-K_d/2}$. Therefore a quartet superfluid phase with dominant
quartet correlations requires $K_d > 2$. The value of $K_d$ has been
computed in QMC using the formula:
\begin{equation}
K_d=\frac{\pi}{4} \lim_{q\rightarrow 0} \frac{N(q)}{q},
\label{kd}
\end{equation}
where the factor $4$ comes from the four spin states. This procedure
has been shown to be very accurate for the spin-1/2 Hubbard
model~\cite{Ejima2005}. For DMRG calculations, $K_d$ can be
independently obtained from the power-law behavior of the quartet
correlations $Q(x)$. Both QMC and DMRG results are shown for example
on the SU(4) invariant line ($V=0$) in Fig.~\ref{Kd_on_SU4.fig}(b).
We find that QMC works better than DMRG in the weak coupling limit and
is in excellent agreement with the perturbative estimate: $K_d^{-2}
= 1 + [V + 3U]/(\sqrt{2}\pi t)$~\cite{Wu2005,Lecheminant2005}. For
larger $|U|$, DMRG is more accurate and we found that $K_d$ saturates
at the value $K_d \simeq 1.6$. Note that the perturbative estimate fails beyond $|U|\geq t$ so that numerical approaches become necessary to estimate $K_d$. 
For $V\ne 0$, $K_d$ also saturates at
strong couplings to values smaller than $2$. We therefore conclude
that, though the quartet correlations are quasi-long ranged, the
dominant instability in the confined phase is a $2k_F$-ADW.

{\it Deconfined Phase --} By allowing $V$ to be sufficiently negative,
one can enter a second phase where the one-particle gap is still
finite (see Fig.~\ref{BCS.fig}(b) for $U=0, V=- 3t$) but the
two-particle gap vanishes.  In this phase, pairing correlations become
algebraic with $P(x) \sim x^{-1/2K_d}$ as shown on
Fig.~\ref{BCS.fig}(a) and $Q(x)$ remains critical with $Q(x) \sim
P(x)^4$ which is the prediction of the low-energy approach. In
contrast to the ADW phase, there is no diverging signal at
$2k_F=\pi/2$ in $N(q)$ (see Fig.~\ref{Kd_on_BCS.fig}(a)). We thus
conclude that there is still a spin-gap and the ${\mathbb{Z}}_2$
symmetry ${\cal U}$ is now spontaneously broken which leads to the
formation of a quasi-long-range BCS phase. In addition, there is also
an ADW instability at $4k_F=\pi$ (see Fig.~\ref{Kd_on_BCS.fig}(a)
where $N(q)$ has a maximum at $q= \pi$) which has a power-law decay
$N(x) \sim \cos(\pi x) x^{- 2 K_d}$. We thus need to compute
numerically $K_d$ to fully characterize the dominant instability of
this phase. As in the previous phase, the Luttinger parameter $K_d$
can be extracted either from Eq.~(\ref{kd}) (QMC) or from pairing
correlations (DMRG). As shown on Fig.~\ref{Kd_on_BCS.fig}(b), both
results are compatible and agree with the perturbative estimate when
$|V|<t$. We find that $K_d > 1/2$ so that the dominant instability in
this phase is the BCS singlet pairing.
\begin{figure}[t]
\includegraphics[width=8cm,clip]{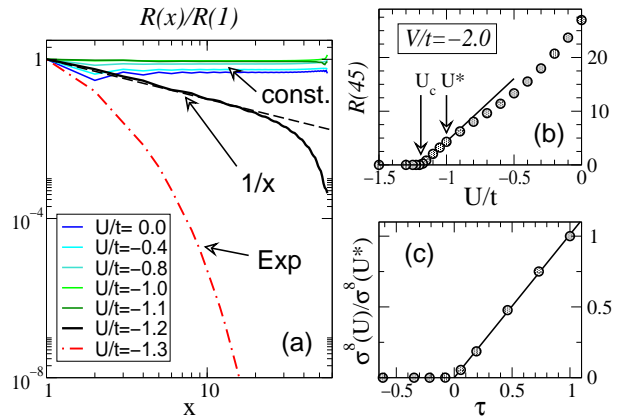}
\caption{(Color online) The BCS-ADW transition from DMRG computations
along the $V=-2t$ line. {\bf(a)} Normalized ratio $R(x)=P(x)^4/Q(x)$
displaying a critical behavior at the transition. {\bf(b)} In the bulk
(at site $x=45$), $R(x)$ is proportional to $U-U_c$ for $U_c\leq U
\leq U^*$ with $U^*=-t$ and $U_c=-1.19t$. {\bf(c)} Plot of
$R(U)=\sigma(U)^8$ vs $\tau = (U-U_c)/(U^*-U_c)$ where $\sigma$ is the
Ising order parameter.}
\label{ising.fig}
\end{figure}
\begin{figure}[t]
\includegraphics[width=8cm,clip]{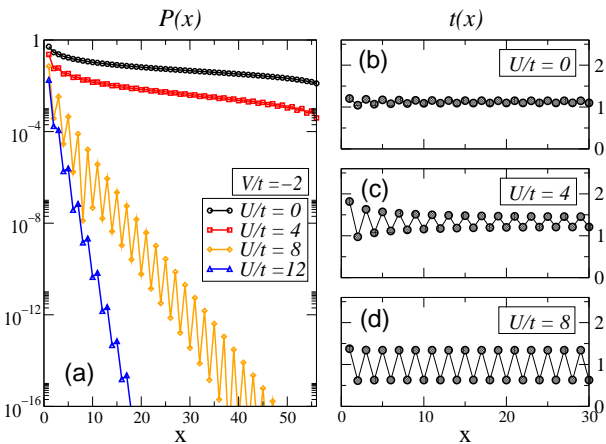}
\caption{(Color online) Mott phase for $V<0$. {\bf (a)}
By increasing $U$, pairing correlations obtained from DMRG change from algebraic to exponential decaying. {\bf (b-d)} The BOW Mott
transition is seen from DMRG computations with the appearance of the
$4k_F$ order parameter $t(x)$ across the transition.}
\label{mott.fig}
\end{figure}

{\it Quantum phase transition --} The striking feature of the phase
diagram for attractive $U,V$ interactions is the change of status of
the ${\mathbb{Z}}_2$ symmetry ${\cal U}$ which is spontaneously broken
(resp. unbroken) in the deconfined (resp. confined) phase. We thus
expect a quantum phase transition in the Ising universality class
between these two phases. In fact it has been shown in
Ref.~\onlinecite{Lecheminant2005} that the order parameter $\sigma(x)$
associated with the ${\cal U}$ symmetry, though being non-local in
terms of the lattice fermions, can be extracted from the long-distance
behavior of the ratio $R(x) = P(x)^4/Q(x)$. In the confined phase
where ${\cal U}$ is unbroken, $\langle \sigma(x) \rangle = 0$ and
$R(x)\sim \langle \sigma(x) \sigma(0) \rangle^4$ decays exponentially
with distance. In the deconfined phase, $\langle \sigma(x) \rangle
= \sigma \neq 0$ and $R(x)\sim \sigma^8$. Finally, it has been found
in Ref.~\onlinecite{Lecheminant2005} that at the transition the ratio
displays an \emph{universal} power-law behavior: $R(x) \sim 1/x$. We
have computed numerically this ratio by DMRG for various parameters to
determine the transition line in Fig.~\ref{phase_diagram.fig}. The
results of Fig.~\ref{ising.fig}(a) clearly show an excellent agreement
with the predictions of the low-energy approach. In particular, we
observe  that $R(x) \sim 1/x$ near the critical point. In the deconfined
phase, $R(x)$ saturates at large distance as it should and is almost
independent of $x$ ($R(x) \sim \sigma(U)^8$) when one enters the
critical regime (for $U_c\leq U\leq U^*$ in
Fig.~\ref{ising.fig}(b)). The plot in Fig.~\ref{ising.fig}(c)
demonstrates that $\sigma(U) \sim (U-U_c)^{1/8}$ in full agreement
with Ising criticality. In this respect, the situation is in sharp
contrast with the $F=1/2$ well-known case where the $2k_F$-ADW and BCS
instabilities coexist for attractive interaction~\cite{bookboso}.

{\it Mott phase --} At quarter-filling, a Mott transition might take
place if $K_d<1/2$ with the formation of a density
gap~\cite{Lecheminant2005,Wu2005}. For the repulsive SU(4) Hubbard
chain ($V=0$), the QMC study of Ref.~\onlinecite{assaraf} found a
transition from a gapless spin-density wave (SDW) to a generalized
Mott SDW with three gapless spin modes (see
Fig.~\ref{phase_diagram.fig}).  For $V<0$, we expect an entirely
different Mott phase due to the presence of a spin gap and the
breaking of the ${\mathbb{Z}}_2$ symmetry ${\cal U}$. In the Mott
region in Fig.~\ref{phase_diagram.fig} (full circles), the BCS singlet pairing becomes short-ranged as shown in Fig.~\ref{mott.fig}(a)
and we find that, as
the density gap opens, the local density almost does not fluctuate and
$N(x) \sim 1$.  In contrast, the local kinetic bond, $t(x) = \langle
\sum_{\alpha} c_{\alpha,x+1}^{\dagger} c_{\alpha,x} + {\rm H.c.}\rangle$,
orders with a $4k_F = \pi$ modulation reminiscent of a {\it doubly}
degenerate ground state as it can be seen in Fig.~\ref{mott.fig}(b-d)
for $V=-2t$.  We therefore conclude on the emergence of a
bond-ordering Mott phase with periodicity two (Mott BOW in
Fig.~\ref{phase_diagram.fig}).

{\it Concluding remarks --} We conclude this letter in emphasizing that
the existence of a quartet superfluid phase where quartet correlations
dominate over the 2k$_F$-ADW instability relies on the non-universal
Luttinger parameter $K_d$. Though $K_d < 2$ at {\it quarter} filling,
we expect that at sufficiently low densities, $K_d$ may become larger
than $2$, which marks the onset of the quartetting phase. 
The formation of this exotic phase will be discussed in a forthcoming paper. 


\acknowledgments We would like to thank K. Totsuka, C. J. Wu, and
S.-C. Zhang for useful discussions. S.~C. and G.~R. thank IDRIS
(Orsay, France) and CALMIP (Toulouse, France) for use of supercomputer
facilities, and Agence Nationale de la Recherche (France) for support.


\begin{thebibliography}{99}

\bibitem{review}
M. Lewenstein {\it et al.},
arXiv: cond-mat/0606771.
\bibitem{ho}
T.-L. Ho, Phys. Rev. Lett. {\bf 81}, 742 (1998).
\bibitem{boson}
T. Ohmi and K. Machida, J. Phys. Soc. Jpn. {\bf 67}, 1822 (1998);
F. Zhou and G. W. Semenoff, Phys. Rev. Lett. {\bf 97}, 180411 (2006);
R. Barnett, A. Turner, and E. Demler {\it ibid.}, 180412 (2006);
R. B. Diener and T.-L. Ho, {\it ibid.} {\bf 96}, 190405 (2006).
\bibitem{spin3}
J. Stenger {\it et al.}, Nature (London) {\bf 396}, 345 (1998);
M. S. Chang {\it et al.}, Nature Phys. {\bf 1}, 111 (2005);
A. Griesmaier {\it et al.}, Phys. Rev. Lett. 
{\bf 94}, 160401 (2005).
\bibitem{nozieres}
P. Schlottmann, J. Phys. Condens. Matter {\bf 6},
1359 (1994);
G. R\"opke, A. Schnell, P. Schuck, and P. Nozi\`eres,
Phys. Rev. Lett. {\bf 80}, 3177 (1998);
B. Dou\c{c}ot and J. Vidal,
{\it ibid.} {\bf 88}, 227005 (2002).
\bibitem{Lecheminant2005}
P. Lecheminant, E. Boulat, and  P. Azaria,
Phys. Rev. Lett. {\bf 95}, 240402 (2005).
\bibitem{Wu2005}
C. J. Wu, Phys. Rev. Lett. {\bf 95}, 266404 (2005).
\bibitem{miyake}
H. Kamei and K. Miyake,
J. Phys. Soc. Jpn. {\bf 74}, 1911 (2005);
A. S. Stepanenko and J. M. F. Gunn,
arXiv: cond-mat/9901317.
\bibitem{zhang} 
C. J. Wu, J. P. Hu, and S.-C. Zhang, Phys. Rev. Lett. {\bf 91}, 186402 (2003);
C. J. Wu, Mod. Phys. Lett. B {\bf 20}, 1707 (2006).
\bibitem{controzzi}
D. Controzzi and A. M. Tsvelik,
Phys. Rev. Lett. {\bf 96}, 097205 (2006).
\bibitem{dmrg} 
S.~R. White, Phys. Rev. Lett. {\bf 69}, 2863 (1992); Phys. Rev. B {\bf
48}, 10345 (1993). U. Schollw\"ock, Rev. Mod. Phys. {\bf 77}, 259
(2005).
\bibitem{Assaad1997} 
F. F. Assaad, M. Imada,
and D. J. Scalapino, Phys. Rev. B {\bf 56}, 15001 (1997).
\bibitem{note} Note that the drop of correlations for the largest distances is due to a finite $M$ effect
and the use of open chains. DMRG is also known to slightly underestimate
correlations (and consequently $K_d$ in this study)~\cite{dmrg}.
\bibitem{Ejima2005} S. Ejima, F. Gebhard,
and S. Nishimoto, Europhys. Lett. {\bf 70}, 492 (2005). 
\bibitem{bookboso}
A. O. Gogolin, A. A. Nersesyan, and A. M. Tsvelik,
{\sl Bosonization and Strongly Correlated Systems}
(Cambridge University Press, Cambridge, England, 1998);
T. Giamarchi, \textit{Quantum Physics in One Dimension}
(Clarendon press, Oxford, UK, 2004).
\bibitem{assaraf} R. Assaraf, P. Azaria, M. Caffarel,
and P. Lecheminant,
Phys. Rev. B {\bf 60}, 2299 (1999).

\end{thebibliography}
\end{document}